\documentclass[aps,prd,
unsortedaddress,superscriptaddress,showpacs,nofootinbib,11pt]{revtex4-2}

\pdfoutput=1

\usepackage{graphicx,color}
\usepackage{relsize}
\usepackage{slashed}
\usepackage[normalem]{ulem}
\usepackage{tabu}
\usepackage{rotating}
\usepackage{bigstrut}
\usepackage{makecell}
\usepackage[dvipsnames]{xcolor}

\usepackage{amsmath}
\usepackage{amssymb,bm}
\usepackage{amsthm}
\usepackage{mathrsfs}
\usepackage{fancyhdr}
\usepackage{array}
\usepackage[all]{xy}
\usepackage{eufrak}
\usepackage{euscript}
\usepackage{enumerate}
\usepackage{slashed}
\usepackage{hyperref}
\usepackage{subfigure} 
\usepackage{mathtools}

\allowdisplaybreaks

\newcommand*{\DStar}{D^{*}}
\newcommand*{\DStarBar}{{\bar D}^{*}}
\newcommand*{\DStarDg}{D^{*\dag}}
\newcommand*{\DStarBarDg}{{\bar D}^{*\dag}}

\newcommand*{\DDg}{D^{\dag}}
\newcommand*{\DBarDg}{{\bar D}^{\dag}}

\usepackage{soul}

\hypersetup{pdftex,colorlinks=true,linkcolor=blue,citecolor=blue,menucolor=black,urlcolor=blue,filecolor=blue}

\newcommand{\itp}{\affiliation{CAS Key Laboratory of Theoretical Physics, Institute of Theoretical Physics,\\
Chinese Academy of Sciences, Beijing 100190, China}}
\newcommand{\ucas}{\affiliation{School of Physical Sciences, University of Chinese Academy of Sciences, Beijing 100049, China}}

\newcommand{\peng}{\affiliation{Peng Huanwu Collaborative Center for Research and Education, Beihang University, Beijing 100191, China}}

\begin{document}

\title{Radiative decays of the spin-$2$ partner of $X(3872)$}

\author{Pan-Pan Shi}
\email{shipanpan@itp.ac.cn}
\itp\ucas

\author{Jorgivan M. Dias}
\email{jorgivan.mdias@itp.ac.cn}
\itp

\author{Feng-Kun Guo}
\email{fkguo@itp.ac.cn}
\itp\ucas\peng 



\begin{abstract}
It has been generally expected that the $X(3872)$ has a spin-2 partner, $X_2$, with quantum numbers $J^{PC}=2^{++}$.
In the hadronic molecular model, its mass was predicted to be below the $D^*\bar D^*$ threshold, and the  new structure reported in the $\gamma \psi(2S)$ invariant mass distribution by the Belle Collaboration with mass $M= (4014.3 \pm 4.0 \pm 1.5)$~MeV and decay width $\Gamma= (4 \pm 11 \pm 6)$~MeV, with a global significance of 2.8 $\sigma$, is a nice candidate for it.
We consider the radiative decay widths for the $X_{2}\to \gamma\psi$ with $\psi=J/\psi, \psi(2S)$ treating the $X_2$ as a $D^*\bar{D}^*$ shallow bound state, and estimate the events of $X_2$ in two-photon collisions that can be collected in the $\gamma J/\psi\to\gamma\ell^+\ell^-$ ($\ell=e,\mu$) final states at Belle. 
Based on the upper limit for the ratio of decay widths of $X(3872)\to \gamma \psi(2S)$ and $X(3872)\to \gamma J/\psi$ measured by BESIII, we predict the similar ratio $\Gamma(X_2\to \gamma \psi(2S))/\Gamma(X_2\to \gamma J/\psi)$ to be smaller than $1.0$. 
We suggest searching for the $X_2$ signal in the $\gamma J/\psi$ invariant mass distribution via two-photon fusions. The results will lead to insights into both the $X(3872)$ and the new structure observed by Belle.
\end{abstract}

\maketitle

\newpage

\section{Introduction}
\label{sec:intro}
The hadron spectroscopy of mesons and baryons 
with heavy quarks (charm and bottom) has been an important 
laboratory in the quest for understanding quantum chromodynamics (QCD) at the 
confinement scale due to the large bulk of experimental 
information accumulated over the last two decades. Many 
new hadronic states observed in the heavy sector seem to 
not fit into the predictions from the potential  quark 
models such as the well-known Godfrey-Isgur quark model~\cite{Godfrey:1985xj}. This fact 
has triggered a debate about the nature of those hadrons, 
and assumptions of different multiquark configurations 
beyond the conventional quark-antiquark/three quarks were 
put forward in order to explain their properties, such as mass, decay width, and the $J^{PC}$ quantum numbers, {\it i.e.}, the 
total angular momentum $J$, parity $P$, and charge conjugation $C$ (see 
the reviews \cite{
Chen:2016qju,Hosaka:2016pey,Esposito:2016noz,Lebed:2016hpi,Ali:2017jda,Olsen:2017bmm,Guo:2017jvc,Albuquerque:2018jkn,Liu:2019zoy,Guo:2019twa,Brambilla:2019esw,Chen:2022asf}).

Among these models, the molecular picture seems to be a 
natural one since the majority of the new 
hadrons are near some hadron-hadron threshold. 
In the charm sector, for instance, the $X(3872)$ state is 
just at the $D\bar{D}^*$ $(\bar{D}D^*)$ threshold, and its properties are suitably described 
considering the $X(3872)$ as a $D \bar{D}^*$ molecular state (for a review focusing on the hadonic molecular model of the $X(3872)$, see Ref.~\cite{Kalashnikova:2018vkv}). 
In fact, the $X(3872)$ was the first among those new hadrons 
observed experimentally by the Belle Collaboration in 2003~\cite{Belle:2003nnu}, 
with $J^{PC}=1^{++}$ quantum numbers determined by the LHCb 
Collaboration a decade later~\cite{LHCb:2013kgk}. It is, to date, the most 
well-studied state, and it is not a surprise that its experimental 
and theoretical information is used as inputs for predictions of 
new hadronic states in the heavy quark sector. In Ref.~\cite{Nieves:2012tt}, 
the authors assumed the $X(3872)$ as a $D\bar{D}^*$ molecule 
and concluded that a $D^*\bar{D}^*$ state should exist as a consequence of the heavy-quark spin symmetry for the system 
under consideration. Specifically, using a contact-range (pionless) 
effective field theory, they claimed that the new state, from 
now on called $X_2$, is the spin-$2$ partner of the $X(3872)$, 
with a similar value for the binding energy and mass of about 
$4012$~MeV. Such a state was first predicted in Ref.~\cite{Tornqvist:1993ng}
long ago with a mass $M=4015$~MeV and later in Refs.~\cite{Molina:2009ct,Nieves:2012tt,Hidalgo-Duque:2012rqv,Sun:2012zzd,Hidalgo-Duque:2013pva,Guo:2013sya,Albaladejo:2013aka,Liang:2010ddf,Swanson:2005tn,Albaladejo:2015dsa,Baru:2016iwj,Cincioglu:2016fkm,Baru:2017fgv,Ortega:2017qmg,Wang:2020dgr,Dong:2021juy,Montana:2022inz} using various phenomenological models.

Recently, the Belle collaboration reported a hint of an isoscalar structure with mass $M=(4014.3 \pm 4.0 \pm 1.5)$
MeV and width $\Gamma_{X_2}=(4\pm 11 \pm 6)$ MeV, seen in 
the $\gamma \psi(2S)$ invariant mass distribution via a 
two-photon process \cite{Belle:2021nuv}.
The global significance is $2.8\sigma$.
This new structure is 
located near the $D^*\bar{D}^*$ threshold which leads us to 
conclude that it is a promising candidate for the $D^*\bar{D}^*$ shallow bound state.\footnote{In Refs.~\cite{Duan:2022upr,Yue:2022gym}, this structure was assumed to be a $D^*\bar D^*$ molecule with $J^{PC}=0^{++}$.} 
In addition, the mass value predicted in Refs.~\cite{Nieves:2012tt,Guo:2013sya} 
is in good agreement with the experimental one reported by Belle~\cite{Belle:2021nuv}, and the measured width well matches the predicted one, of the order of a few MeV, in Ref.~\cite{Albaladejo:2015dsa} despite that there is a sizeable uncertainty in the theoretical predictions~\cite{Albaladejo:2015dsa,Baru:2016iwj}. Thus, this narrow structure could be a hint for the $X_2$ state, supporting the theoretical 
predictions in Refs.~\cite{Nieves:2012tt,Tornqvist:1993ng,Guo:2013sya}.

Alternatively, by looking at the spectra of tetraquarks, there 
also exists spin partners of $1^{++}$ states that reproduce the 
$2^{++}$ quantum numbers of $X_2$ as well as its mass~\cite{Maiani:2014aja,Wu:2018xdi,Shi:2021jyr,Giron:2021sla}. 
Although the authors of these works were not particularly aiming at the $X_2$ state, 
it is still possible to assign the results to such a structure. 
On the other hand, a $2^{++}$ tensor state with a similar mass
could also be described as a conventional $2P$ charmonium state~\cite{Godfrey:1985xj,Li:2009ad}; in this case, the $\chi_{c2}(3930)$~\cite{ParticleDataGroup:2022pth} would be an exotic meson.

One way to disentangle those different multiquark configurations 
from the molecular point of view is to check the mass splitting
between the $2^{++}$ and $1^{++}$ states. In Refs.~\cite{Nieves:2012tt,Guo:2013sya}, 
the corresponding mass splitting is approximately equal to that 
between the vector and pseudoscalar charmed mesons, that is
\begin{equation}
m_{X_2}-m_X \sim m_{D^*} - m_D \sim 140 ~\textrm{MeV}\, ,
\label{msplit}
\end{equation}
with $m_D (m_{D^*})$ the pseudoscalar (vector) charmed meson 
mass. On the other hand, within the tetraquark approach, for 
instance, in Ref.~\cite{Maiani:2014aja} the $2^{++}-1^{++}$ is 
about $80$~MeV, which is smaller than the difference given in 
Eq.~\eqref{msplit}. A similar conclusion is found for the difference 
between the first radially excited charmonia $2^{++}$ and $1^{++}$ 
by looking at the results for both the Godfrey-Isgur quark model~\cite{Godfrey:1985xj} 
and the one using a screened potential~\cite{Li:2009ad}, which are about $30$~MeV and $40$~MeV, 
respectively.

Even though only the Belle experiment has reported a signal relevant to the spin-$2$ partner of the $X(3872)$, such a structure 
can also be searched for in other ongoing and future experiments, e.g., BESIII and its upgrade, 
LHCb, and PANDA. Belle II also has plans to search 
for the $X_2$ state soon. In line with the current and upcoming 
experiments that will provide more information about 
such a structure, it is crucial to extend the theoretical studies 
surveying the $X_2$ system. In other words, we should further 
explore the $2^{++}$ tensor state to help discriminate 
the various multiquark models used to describe the $X_2$ structure. 

The decays of a $2^{++}$ tensor structure have been studied in 
Refs.~\cite{Barnes:2003vb,Barnes:2005pb}. In particular, considering 
that state as the first radial excitation of the $P$-wave $\chi_{c2}$ ($2^3P_2$) 
charmonium, the quark model adopted in Ref.~\cite{Barnes:2003vb,Barnes:2005pb} 
provides width estimates for the $X_2$ decay to charmed mesons 
around tens of MeV. Moreover, the hadronic decays of the
$D^*\bar{D}^*$ $S$-wave hadronic molecule, into $D\bar{D}$
and $D\bar{D}^*$ meson pairs were estimated to be of the order of a few MeV in Ref.~\cite{Albaladejo:2015dsa} and can be as large as 50~MeV in Ref.~\cite{Baru:2016iwj}.

Furthermore, the $X_2\to \gamma 
D\bar{D}^*$ decay width was also calculated in Ref.~\cite{Albaladejo:2015dsa} to be of the keV order. 
In contrast to the hadronic decays that, according to Ref.~\cite{Albaladejo:2015dsa}, 
has a strong dependence on the ultraviolet (UV) form factors and therefore are 
sensitive to the short-distance details, 
radiative decays into $\gamma 
D\bar{D}^*$ are more sensitive to the long-distance structure 
of the resonance. Thus, as argued in Ref.~\cite{Albaladejo:2015dsa}, 
one can extract valuable information about 
the $X_2$ wave function, as well as about $D\bar D^*$ interactions, by surveying such decays. It is not difficult to understand this 
feature.
In the $(D^*\bar{D}^*)\to \gamma D D^*$ process, the final 
state receives leading contribution from the one-body transition $D^*\to D\gamma$, which has no direct relation to the two-body interaction accounting for the short-distance part of the $X_2$ state. 
Therefore, the long-range 
structure of $X_2$ that determines its coupling to $D^*\bar D^*$ has an essential role in its radiative decay into $\gamma 
D\bar{D}^*$.

Additional interesting decay modes of the $X_2$ which have not been explored before include the
radiative decays into $\gamma \psi(2S)$ and $\gamma J/\psi$ 
channels. 
Such a study may help to discriminate the $X_2$ nature   from the $c\bar c$ meson $\chi_{c2}(2P)$ possibility. 
As discussed in, e.g.,
Ref.~\cite{Guo:2014taa}, the radiative decay matrix element 
is proportional to the overlap between the wave functions 
corresponding to the initial and final states. Specifically, for transitions between two charmonia, 
that overlap is influenced by the position of the nodes of the 
wave function. Hence, the one-node wave function for the $\psi(2S)$ 
state has an overlap with the $\chi_{c2}(2P)$ larger than that for the $J/\psi$ one, which 
is nodeless, such that the following ratio
\begin{align}
    R_{X_2}\equiv\frac{\text{Br}\left(X_2\to \gamma\psi(2S)\right)}{\text{Br}\left(X_2\to \gamma J/\psi\right)},
\end{align}
should be much larger than the one if the initial particle is the $\chi_{c2}(2P)$ charmonium. Table~\ref{Tab:charmonium} shows some 
results for $R_{X_2}$ obtained in different quark models~\cite{Barnes:2003vb,Barnes:2005pb}. 
Therefore, to confront these results, we evaluate the $X_2$ 
radiative decays into $\gamma \psi(2S)$ and $\gamma J/\psi$, assuming 
that the $X_2$ resonance is a $D^*\bar{D}^*$ molecular partner of the 
$X(3872)$ state, predicted in Refs.~\cite{Nieves:2012tt,Guo:2013sya} 
according to heavy quark spin symmetry (HQSS).

\begin{table}[tb]
    \caption{\label{Tab:charmonium} Some results for the radiative 
    decays of the $2^3P_2$ charmonium calculated with quark model. 
    $\Gamma_{\psi}$ denotes the decay width for $2^3P_2\to 
    \gamma\psi$ with $\psi=J/\psi, \psi(2S)$.
    }
    \renewcommand{\arraystretch}{1.2}
    \begin{tabular*}{\columnwidth}{@{\extracolsep\fill}lcccc}
    \hline\hline                                           
                & $\Gamma_{J/\psi}$ [keV]   & $\Gamma_{\psi(2S)}$ [keV] & $R_{X_2}$  \\[5pt]        
    \hline
    Ref. \cite{Barnes:2003vb} & 53  & 207   & 3.9 \\[5pt]
    Ref. \cite{Barnes:2005pb} & 81  & 304   & 3.8 \\[5pt]
    \hline\hline
    \end{tabular*}
    \end{table}
In order to calculate those radiative decay widths, we employ the 
couplings of the $\psi$ mesons to charmed mesons respecting HQSS and the magnetic and electric couplings 
of charmed mesons and a photon. The nonrelativistic effective 
field theory is applied to depict the coupling of the $X_2$ to 
$D^*\bar{D}^*$, which is related to the binding energy of $X_2$.
After calculating the radiative decay widths for $X_2\to \psi\gamma$ 
with $\psi=J/\psi, \psi(2S)$, the upper limit of the ratio $R_{X_2}$ 
is predicted with the use of the experimental result reported by 
the BESIII collaboration $R_{X(3872)}< 0.59$ at $90 \%$ confidence 
level~\cite{BESIII:2020nbj}, and the results for the radiative decays of $X(3872)$ in the hadronic molecular picture~\cite{Guo:2014taa}. Taking into account the signal yield of $X_2$ 
in the $\gamma\psi(2S)$ invariant mass distribution measured by Belle~\cite{Belle:2021nuv}, 
and the predicted upper bound of $R_{X_2}$, we also estimate the lower 
limit of the signal yield of $X_2$ in the $\gamma J/\psi$ mode via the two-photon process at Belle, with $J/\psi$ reconstructed in lepton-antilepton pairs. 

The structure of the paper is as follows. In Section~\ref{Sec:Formalism}, 
we discuss the interaction Lagrangians and the main parameters 
used as well as the relevant Feynman diagrams contributing to the 
decay process $X_2\to \gamma\psi$.
The radiative decay ratio $R_{X_2}$ and the signal yield of $X_2$ 
in the $\gamma J/\psi$ mode are predicted in 
Section~\ref{Sec:Result}. A summary is given in Section~\ref{Sec:Summary}. 
Finally, in Appendix~\ref{Sec:X_3872}, we provide an update for the 
results corresponding to the radiative decay widths for $X(3872)\to 
\gamma\psi$ discussed in Ref.~\cite{Guo:2014taa}.

\section{Formalism}
\label{Sec:Formalism}

\subsection{The Lagrangian and vertices} 

\begin{figure}[tbp]
    \subfigure[ ]{
    \includegraphics[width=0.3\columnwidth]{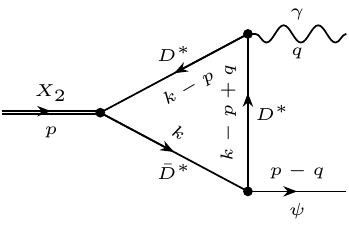} } {\hglue 0.4cm}
    \subfigure[ ]{
    \includegraphics[width=0.3\columnwidth]{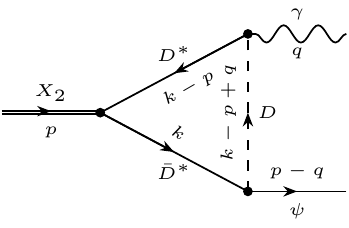}}{\hglue 0.4cm}
    \subfigure[ ]{
    \includegraphics[width=0.3\columnwidth]{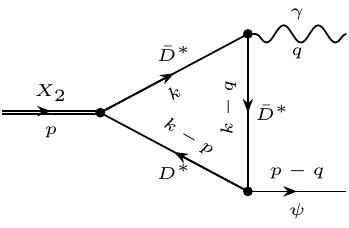}}\\[10pt]
    \subfigure[ ]{
    \includegraphics[width=0.3\columnwidth]{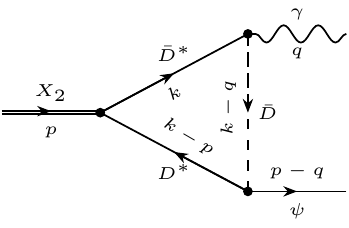} } {\hglue 0.4cm}
    \subfigure[ ]{
    \includegraphics[width=0.3\columnwidth]{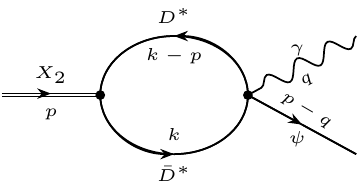}}{\hglue 0.4cm}
    \subfigure[ ]{
    \includegraphics[width=0.18\columnwidth]{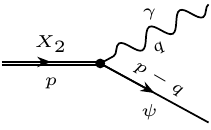}}{\hglue 0.4cm}
    \caption{Feynman diagrams for $X_2\rightarrow \gamma \psi$ ($\psi=J/\psi,\psi(2S)$).}
    \label{Fig:decay_X_2}
\end{figure}

As discussed in Refs.~\cite{Guo:2009wr,Guo:2010ak,Guo:2010ca} hadron 
loops play an important role in certain hadron transitions. For pure hadronic molecules, 
those loops are the leading order contribution to the corresponding transition 
amplitudes due to the large coupling of the molecule to its constituents. 
In our case, the $X_2$ radiative decays under consideration 
proceeds through the loops depicted in Fig.~\ref{Fig:decay_X_2}.
In order to evaluate each diagram displayed 
in Fig.~\ref{Fig:decay_X_2}, we need first to define the 
interaction Lagrangian that describes all the vertices involved in 
such loops. We start by the $X_2\, D^* \bar{D}^*$ interaction 
vertex that is described by the following Lagrangian
\begin{align}
    {\cal L}_{X_2}=\chi^0_{\text{nr}} X_2{}_{\mu\nu}^{\dag}D^{*0\mu}{\bar D}^{*\nu} 
    +\chi^c_{\text{nr}} X_2{}_{\mu\nu}^{\dag}D^{*+\mu}D^{*-\nu}+\text {h.c.},
    \label{x2int}
\end{align}
where $\chi_\text{nr}^0$ and $\chi_\text{nr}^c$ are the $X_2$ couplings to the neutral (0)
and charged $(c)$ charmed mesons, while the subscript ``nr" stands 
for ``nonrelativistic". As we know, there is a slight difference between 
the neutral and charged meson masses that leads to an isospin-breaking effect.
However, according to  Refs.~\cite{Nieves:2012tt,Albaladejo:2015dsa}, 
this effect is small so that
the couplings $\chi_{\rm nr}^0$ and $\chi_{\rm nr}^c$ are approximately the same. In addition, the relative size 
of $X_2$ is much smaller compared to the Bohr radius of a ground-state 
hadronic atom, made out of the charged $D^{*+}$ and $D^{* -}$ mesons, 
so that the electromagnetic effects can be ignored in such a scenario. 
Therefore, we follow Ref.~\cite{Guo:2014taa} and set $\chi_\text{nr}^0=\chi_\text{nr}^c = \chi_\text{nr}$. The $X_2 D^* \bar{D}^*$ vertex is then
\begin{align}
    \Gamma_{\mu\nu\alpha\beta}^{X_2}=i\chi_\text{nr}\,g_{\mu\alpha}\,g_{\nu\beta}\,.
    \label{Eq:coupling_X2}
\end{align}

Next, the interaction vertex between the $\psi$ and charmed $D^{(*)}$ and 
$\bar{D}^{(*)}$ mesons can be extracted from 
\begin{align}
    {\cal L}_{\psi}=&\,g_{2} \psi_{\mu}\left(\DStarBarDg{}^{\nu} \overleftrightarrow{\partial}_{\nu}
    \DStarDg{}^{\mu}
    +\DStarBarDg{}^{\mu} \overleftrightarrow{\partial}_{\nu}
    \DStarDg{}^{\nu}-
    \DStarBarDg{}^{\nu} \overleftrightarrow{\partial}^{\mu}
    \DStarDg{}_{\nu}\right)
    -g_{2} \psi_{\mu}\DBarDg \overleftrightarrow{\partial}^{\mu}
    \DDg \nonumber\\[3pt]
    &-ig_{2} \epsilon^{\mu\nu\alpha\beta}\psi_{\mu}v_{\alpha}\left(\DStarBarDg{}_{\nu}
    \overleftrightarrow{\partial}_{\beta}\DDg
    -\DBarDg
    \overleftrightarrow{\partial}_{\beta}\DStarDg{}_{\nu}\right)+ \text{h.c.},
    \label{Eq:Lagrangian_DDpsi}
\end{align}
encoding HQSS~\cite{Jenkins:1992nb,Colangelo:2003sa}. In 
Eq.~\eqref{Eq:Lagrangian_DDpsi}, $v_{\alpha}$ is the four-velocity of the charmed meson. 
By defining the four-momentum as $p_{\alpha}=m_{D^{(*)}} v_{\alpha}+ k_{\alpha}$, 
with $k_{\alpha}$ a residual momentum of $\mathcal{O}(\Lambda_\text{QCD})$, 
and recalling that $v^2=1$, 
we can write $v_{\alpha}$ as 
\begin{equation}
 v_{\alpha}= \frac{p_{\alpha}}{m_{D^{(*)}}}-{\cal O}\left(\frac{k_{\alpha}}{m_{D^{(*)}}}\right)\,,
\end{equation} 
where $m_{D^{(*)}}$ is the charmed meson mass. Furthermore, we 
can write the coupling constant $g_2$ in terms of the relativistic couplings 
$g_{D \bar{D}}, g_{D \bar{D}^*}$, and $g_{D^* \bar{D}^*}$ as given 
in Refs.~\cite{Colangelo:2003sa,Guo:2010ak,Guo:2014taa}, that is
\begin{align}
    g_{{\bar D}D}=g_2m_D\sqrt{m_{\psi}}, \quad
    g_{\DStarBar D}=2g_2\sqrt{\frac{m_Dm_{\psi}}{m_{D^*}}}, \quad g_{\DStarBar\DStar}=g_2m_{D^*}\sqrt{m_{\psi}},
    \label{Eq:Coupling_g2}
\end{align}
where $m_{\psi}$ is the mass of the $\psi$ meson. From Eq.~\eqref{Eq:Lagrangian_DDpsi} 
we extract the interaction vertices $\psi_{\mu}(p) \to 
\bar{D}^*_{\nu}(k_1) D(-k_2)$ and $\psi_{\mu}(p) 
\to \bar{D}^*_{\alpha}(k_1) D^*_{\beta}(-k_2)$, which are
\begin{align}
\Gamma^{(\DStarBar D)}_{\mu\nu}&=-i\frac{2g_2}{m_{D^*}}\epsilon_{\mu\nu\alpha\beta}\,k_1^{\alpha}\,k_2^{\beta},\\[3pt]
\Gamma^{(\DStarBar\DStar)}_{\mu\alpha\beta}&=g_2\left[ (k_1+k_2)_{\alpha}\,g_{\mu\beta}+(k_1+k_2)_{\beta}\,g_{\mu\alpha}-(k_1+k_2)_{\mu}\,g_{\alpha\beta}\right].\label{Eq:vertex_D1_D1_psi}
\end{align}

Now, we move on to the interaction between the charmed mesons 
and the photon. In this case, we have two couplings corresponding to 
the electric and magnetic interactions. The former is 
obtained by gauging the kinetic term associated with the 
charged $D^{(*)}$ mesons, which is
\begin{align}
    {\cal L}_e =&\, \partial_{\mu}\DDg\partial^{\mu}D-m_{D}^2\DDg D+ie\text{Q}_{D}A_{\mu}\left(\partial^{\mu}\DDg D-\DDg\partial^{\mu}D\right)+e^2\text{Q}_D^2A_{\mu}A^{\mu}\DDg D-\frac{1}{2}\DStarDg_{\mu\nu}\DStar{}^{\mu\nu}\nonumber\\
    &+m_{\DStar}^2\DStarDg_{\mu}\DStar{}^{\mu}
    +ie\text{Q}_{D^*}A_{\mu}\left(\DStarDg_{\nu}~\overleftrightarrow{\partial}^{\mu} \DStar{}^{\nu}+\partial_{\nu}\DStarDg{}^{\mu} \DStar{}^{\nu}-\DStarDg{}^{\nu} \partial_{\nu} \DStar{}^{\mu}\right)\nonumber\\
    &-e^2\text{Q}_{\DStar}^2\left(A_{\mu}A^{\mu}\DStarDg_{\nu}\DStar{}^{\nu}-A_{\mu}A_{\nu}\DStarDg{}^{\nu}\DStar{}^{\mu}\right),
    \label{Eq:electric_couple}
\end{align}
with $e\text{Q}_{D^{(*)}}$ standing for the electric charge of the heavy $D^{(*)}$ 
meson, and $\DStar_{\mu\nu}=\partial_{\mu}\DStar_{\nu}-\partial_{\nu}\DStar_{\mu}$. 
The $\DStar_{\mu}{}^{\pm}(k_1)\to \DStar_{\alpha}{}^{\pm}(k_2)\gamma_{\beta}(q)$ vertex reads
\begin{align}
 \Gamma^{(e)}_{\mu\alpha\beta}&=ie|\text{Q}_{D^{*}}|\left[ (k_1+k_2)_{\beta}g_{\mu\alpha}-k_1{}_{\alpha}g_{\mu\beta}-k_2{}_{\mu}g_{\alpha\beta}\right].
\label{Eq:vertex_D1_D1_Gamma}
\end{align}
Note that this vertex satisfies the Ward-Takahashi identity, as discussed in Ref.~\cite{Guo:2014taa}. Besides, after gauging the vertex of Eq. \eqref{Eq:vertex_D1_D1_psi}, a four-point vertex for $\psi_{\mu}(p)\gamma_{\nu}(q) \to D^{*-}_{\alpha}(k_1)D^{*+}_{\beta}(-k_2)$, see the diagram in Fig. \ref{Fig:decay_X_2}~(e), reads
\begin{align}
\Gamma^{(e)}_{\mu\nu\alpha\beta}=2e|\text{Q}_{D^{*}}|g_2\left(g_{\mu\beta}g_{\nu\alpha}+g_{\mu\alpha}g_{\nu\beta}-g_{\mu\nu}g_{\alpha\beta}\right).
\label{Eq:vertex_D1_D1_Gamma_psi}
\end{align}

On the other hand, the magnetic vertices are extracted from the Lagrangian~\cite{Amundson:1992yp}
\begin{align}
    {\cal L}_m
    =&-ieF^{\mu\nu}\left(\DStarDg_{\mu}\DStar_{\nu}-\DStarDg_{\nu}\DStar_{\mu}\right)\left(\frac{Q\beta'}{2}-\frac{Q'}{2m_c}\right)+e\epsilon^{\mu\nu\alpha\beta}F_{\mu\nu}v_{\alpha}\left(\DDg\DStar_{\beta}+\DStarDg_{\beta}D\right)\left(\frac{Q\beta'}{2}+\frac{Q'}{2m_c}\right)\nonumber\\[3pt]
    &+ieF^{\mu\nu}\left(\DStarBarDg_{\mu}\DStarBar_{\nu}-\DStarBarDg_{\nu}\DStarBar_{\mu}\right)\left(\frac{Q\beta'}{2}-\frac{Q'}{2m_c}\right)
    +e\epsilon^{\mu\nu\alpha\beta}F_{\mu\nu}v_{\alpha}\left(\DBarDg\DStarBar_{\beta}+\DStarBarDg_{\beta}{\bar D}\right)\left(\frac{Q\beta'}{2}+\frac{Q'}{2m_c}\right),
  \label{Eq:magnitic_couple}
\end{align}
where $Q=\text{Diag}(2/3,-1/3)$ is the light quark charge matrix, and $Q'=2/3$ 
corresponds to the charge of the charm quark, while the $F_{\mu\nu}$ stands 
for the electromagnetic field tensor. In addition, $m_c$ is the charm quark mass, 
and the parameter $\beta^{\prime}$ is discussed in Ref.~\cite{Hu:2005gf}. From 
Eq.~\eqref{Eq:magnitic_couple} the vertices $\DStar_{\mu}(k_1)\to 
\gamma_{\nu}(q) D(k_2)$, $\DStar_{\mu}(k_1)\to \gamma_{\beta}(q)\DStar_{\alpha}(k_2)$, 
$\DStarBar_{\mu}(k_1)\to \gamma_{\nu}(q){\bar D}(k_2)$ and $\DStarBar_{\mu}(k_1)\to 
\gamma_{\beta}(q)\DStarBar_{\alpha}(k_2)$ are
\begin{align}
    \Gamma^{(m)\DStar D\gamma}_{\mu\nu}&=e\epsilon_{\mu\nu\alpha\beta}v_{\alpha}q_{\beta}\left(\beta'Q+\frac{Q'}{m_c}\right),\label{Eq:vertex_D1_D_gamma_M}\\[3pt]
    \Gamma^{(m)\DStar\DStar\gamma}_{\mu\alpha\beta}&=ie(q_{\alpha}g_{\mu\beta}-q_{\mu}g_{\alpha\beta})\left(\beta' Q-\frac{Q'}{m_c}\right),\label{Eq:vertex_D1_D1_gamma_M}\\[3pt]
    \Gamma^{(m)\DStarBar {\bar D}\gamma}_{\mu\nu}&=e\epsilon_{\mu\nu\alpha\beta}v_{\alpha}q_{\beta}\left(\beta' Q+\frac{Q'}{m_c}\right),\label{Eq:vertex_D1B_DB_gamma_M}\\[3pt]
    \Gamma^{(m)\DStarBar\DStarBar\gamma}_{\mu\alpha\beta}&=-ie(q_{\alpha}g_{\mu\beta}-q_{\mu}g_{\alpha\beta})\left(\beta' Q-\frac{ Q'}{m_c}\right).\label{Eq:vertex_D1B_D1B_gamma_M}
\end{align}
As in Eq.~\eqref{Eq:vertex_D1_D1_Gamma}, these vertices also satisfy the appropriate Ward identity.

\subsection{The amplitude $X_2\rightarrow \gamma \psi$}
Once we have all the vertices in the loop diagrams in Fig.~\ref{Fig:decay_X_2}, 
we are able to write the amplitude for the $X_2 \to \gamma \psi$ decay. Namely,
\begin{equation}
    i{\cal M}=ie\,\chi_{\text{nr}}\, g_2\, \epsilon^*{}^{\mu\nu}(X_2)\epsilon^{\beta}(\gamma)\epsilon^{\sigma}(\psi)\, {\cal M}_{\mu\nu\beta\sigma},
    \label{Eq:amp_X_decay}
\end{equation}
with $\varepsilon^*{}^{\mu \nu}(X_2), \varepsilon^{\beta}(\gamma)$, 
and $\varepsilon^{\sigma}(\psi)$ the polarization vectors for the 
$X_2$ state, the photon, and $\psi$ mesons ($\psi^{\prime}$, and $J/\psi$), 
respectively. The tensor structure ${\cal M}_{\mu\nu\beta\sigma}$ in Eq.~\eqref{Eq:amp_X_decay} 
encodes the contributions of all the diagrams in Fig.~\ref{Fig:decay_X_2}, and it is written as
\begin{align}
    {\cal M}_{\mu\nu\beta\sigma}=&\,\sqrt{m_{X_2}m_{\psi}}\int\frac{d^4k}{(2\pi)^4}S_{\nu}^{\rho}(k)S_{\mu}^{\alpha}(k-p)\nonumber\\[3pt]
    &\left(J_{\alpha\rho\beta\sigma}^{(a)m}(k)+J_{\alpha\rho\beta\sigma}^{(a)e}(k)+J_{\alpha\rho\beta\sigma}^{(b)m}(k)+J_{\alpha\rho\beta\sigma}^{(c)m}(k)+J_{\alpha\rho\beta\sigma}^{(c)e}(k)+J_{\alpha\rho\beta\sigma}^{(d)m}(k)+J_{\alpha\rho\beta\sigma}^{(e)e}(k)\right),
    \label{Eq:amp_total}
\end{align}
where the superscripts (a)-(e) match the labels of each individual diagram in 
Fig.~\ref{Fig:decay_X_2}. Yet, the contributions in Eq.~\eqref{Eq:amp_total} 
with the labels $m$ and $e$ corresponds to the 
magnetic and electric couplings, respectively. Explicitly, each contribution in 
Eq.~\eqref{Eq:amp_total} reads
\begin{align}
    J_{\alpha\rho\beta\sigma}^{(a)m}(k)=&\frac{i}{3}m_{D^*}^3S^{\xi\gamma}(k-p+q) \left[ (2k-p+q)_{\sigma}g_{\rho\xi} -(2k-p+q)_{\rho}g_{\sigma\xi} -(2k-p+q)_{\xi}g_{\rho\sigma} \right]\nonumber\\[3pt]
    &(q_{\alpha}g_{\gamma\beta}-q_{\gamma}g_{\alpha\beta})\left(\beta'-\frac{4}{m_c}\right),\label{Eq:amp_a_m}\\[3pt]
    J_{\alpha\rho\beta\sigma}^{(a)e}(k)=&im_{D^*}^2 S^{\xi\gamma}(k-p+q) \left[ (2k-p+q)_{\sigma}g_{\rho\xi} -(2k-p+q)_{\rho}g_{\sigma\xi} -(2k-p+q)_{\xi}g_{\rho\sigma} \right]\nonumber\\[3pt]
    &\left[ (2k-2p+q)_{\beta}g_{\alpha\gamma} -(k-p+q)_{\alpha}g_{\gamma\beta} -(k-p)_{\gamma}g_{\alpha\beta} \right],\label{Eq:amp_a_e}\\[3pt]
    J_{\alpha\rho\beta\sigma}^{(b)m}(k)=&-\,\frac{2i}{3}m_{D^*}S(k-p+q) \epsilon_{\sigma\rho\gamma\delta}k^{\gamma}(k-p+q)^{\delta}\epsilon_{\alpha\beta\xi\eta}(k-p)^{\xi}q^{\eta}
    \left(\beta'+\frac{4}{m_c}\right),\label{Eq:amp_b_m}\\[3pt]
    J_{\alpha\rho\beta\sigma}^{(c)m}(k)=&-\,\frac{i}{3}m_{D^*}^3S^{\xi\gamma}(k-q) \left[ (2k-p-q)_{\sigma}g_{\alpha\xi} -(2k-p-q)_{\xi}g_{\sigma\alpha} -(2k-p-q)_{\alpha}g_{\sigma\xi} \right]\nonumber\\[3pt]
    &(q_{\gamma}g_{\rho\beta}-q_{\rho}g_{\beta\gamma})\left(\beta'-\frac{4}{m_c}\right),\label{Eq:amp_c_m}\\[3pt]
    J_{\alpha\rho\beta\sigma}^{(c)e}(k)=&im_{D^*}^2S^{\xi\gamma}(k-q) \left[ (2k-p-q)_{\sigma}g_{\alpha\xi} -(2k-p-q)_{\xi}g_{\sigma\alpha} -(2k-p-q)_{\alpha}g_{\sigma\xi} \right]\nonumber\\[3pt]
    &\left[ (2k-q)_{\beta}g_{\rho\gamma} -k_{\gamma}g_{\rho\beta} -(k-q)_{\rho}g_{\gamma\beta} \right],\label{Eq:amp_c_e}\\[3pt]
    J_{\alpha\rho\beta\sigma}^{(d)m}(k)=&\frac{2i}{3}m_{D}S(k-q) \epsilon_{\rho\beta\xi\eta}k^{\xi}q^{\eta}\epsilon_{\sigma\alpha\gamma\delta}(k-q)^{\gamma}(k-p)^{\delta}
    \left(\beta'+\frac{4}{m_c}\right),\label{Eq:amp_d_m}\\[3pt]
    J_{\alpha\rho\beta\sigma}^{(e)}(k)=&\,2m_{D^*}^2 \left(g_{\alpha\sigma}g_{\beta\rho}+ g_{\sigma\rho}g_{\alpha\beta}-g_{\alpha\rho}g_{\beta\sigma}\right)\, ,
    \label{Eq:amp_e}
\end{align}
with $S$ and $S_{\mu\nu}$ the propagator for the heavy fields $D$ and $D^*$, respectively, given by
\begin{align}
    S(\tilde{p})&=\frac{i}{\tilde{p}^2-m_D^2+i\epsilon},\nonumber\\[3pt]
    S_{\mu\nu}(\tilde{p})&=\frac{i}{\tilde{p}^2-m_{D^*}+i\epsilon}\left(-g_{\mu\nu}+\frac{\tilde{p}_{\mu}\tilde{p}_{\nu}}{m_{D^*}^2}\right).
\end{align}
The factor $\sqrt{m_{X_2}\,m_{\psi}}$ in Eq.~\eqref{Eq:amp_total} accounts for the 
normalization of the heavy meson fields.\footnote{{Except for the electric coupling 
in Eq.~\eqref{Eq:vertex_D1_D1_Gamma}, we use the nonrelativistic normalization for 
the heavy mesons (including the charmonium, charmed mesons, and $X_2$), which 
differs from the traditional relativistic normalization by a factor $\sqrt{m_H}$.}} 
It is worth noticing that, as a consistency check, the loop amplitude in 
Eq.~\eqref{Eq:amp_total} is gauge invariant, as we must expect.

The loop amplitude defined in Eq.~\eqref{Eq:amp_total} is a UV divergent 
integral. In order to have a well-defined amplitude from which we can get 
consistent results, we add a $X_2\gamma \psi$ counterterm amplitude like the case for the $X(3872)\to \gamma\psi$~\cite{Guo:2014taa}, 
depicted in Fig.~\ref{Fig:decay_X_2}(f), to the one given in Eq.~\eqref{Eq:amp_total}. 
Specifically, it is 
\begin{align}
    i{\cal M}^{\text{cont}}=&\, i\lambda_1\epsilon^*_{\mu\nu}(X_2)\epsilon^{\mu}(\gamma)\epsilon^{\nu}(\psi)
    +i\lambda_2q^{\mu}q^{\nu}\epsilon^*_{\mu\nu}(X_2)p\cdot\epsilon(\gamma)q\cdot\epsilon(\psi)\nonumber\\[3pt]
    &+i\lambda_3 q^{\mu}\epsilon^{\nu}(\psi)\epsilon^*_{\mu\nu}p\cdot\epsilon(\gamma)
    +i\lambda_4 q^{\mu}\epsilon^{\nu}(\gamma)\epsilon^*_{\mu\nu}p\cdot\epsilon(\psi)
    +i\lambda_5 q^{\mu}q^{\nu}\epsilon^*_{\mu\nu}\epsilon(\gamma)\cdot\epsilon(\psi),
    \label{Eq:amp_conuterterm}
\end{align}
which is straightforward to see its manifest gauge invariance. These terms are defined
such that  $\lambda_r$ ($r=1,\,.\,.\,.,5$), which is subject to renormalization, 
absorbs the UV divergence from the loops in Eq.~\eqref{Eq:amp_total}, 
as done in Ref.~\cite{Guo:2014taa}. On the one hand, the counterterm defined in 
Ref.~\cite{Guo:2014taa} has only one parameter $\lambda$; on the other hand, 
in our case, Eq.~\eqref{Eq:amp_conuterterm} has five terms with each one having 
its strength $\lambda_r$. However, for our purposes, we do not consider 
relations among them.

The $X_2\to \gamma \psi$ two-body decay width
is given by the following formula
\begin{align}
    \Gamma_{X_2}=&\,e^2\chi_{\text{nr}}^2 g_2^2\sum_{\text{polarizations}}\frac{1}{5}\frac{q}{8\pi m_{X_2}^2}\left| \epsilon^{*}{}^{\mu\nu}(X_2)\epsilon^{\beta}(\gamma)\epsilon^{\sigma}(\psi) {\cal M}_{\mu\nu\beta\sigma}\right|^2\nonumber\\[3pt]
    =&-\frac{e^2\chi_{\text{nr}}^2 g_2^2~{q}}{40\pi m_{X_2}^2}{\cal M}_{\mu\nu\beta\sigma}\,{\cal M}_{\mu'\nu'\beta'\sigma'}^*\bar{g}^{\sigma\sigma'}(p-q,m_{\psi})g^{\beta\beta'}P^{(2)}{}^{\mu\nu\mu'\nu'}(p,m_{X_2}),
    \label{Eq:X2decay}
\end{align}
where $q=|\Vec{q\,}|$ stands for the momentum of the final states ($\gamma$ or $\psi$) 
in the center-of-mass frame, 
\begin{align}
    q=\frac{m_{X_2}^2-m_{\psi}^2}{2m_{X_2}}.
\end{align}
Furthermore, $P^{(2)}_{\mu\nu\mu'\nu'}(p,m_{X_2})$ in Eq.~\eqref{Eq:X2decay} 
is the projection operator corresponding to the summation over the polarizations 
of $X_2$, which is given by~\cite{Chung:1971ri}
\begin{align}
    P^{(2)}_{\mu\nu\mu'\nu'}(p,m_{X_2})=&\sum_{\text{polarizations}}\epsilon_{\mu\nu}(p,m_{X_2})\epsilon_{\mu'\nu'}(p,m_{X_2})\nonumber\\[3pt]
    =&\,\frac{1}{2}\left[\bar{g}_{\mu\mu'}(p,m_{X_2})\bar{g}_{\nu\nu'}(p,m_{X_2})+\bar{g}_{\mu\nu'}(p,m_{X_2})\bar{g}_{\nu\mu'}(p,m_{X_2})\right]\nonumber\\
    &-\frac{1}{3}\bar{g}_{\mu\nu}(p,m_{X_2})\bar{g}_{\mu'\nu'}(p,m_{X_2}),
\end{align}
where $\bar{g}_{\alpha\beta}$ denotes the summation over the polarization vectors 
\begin{align}
        \bar{g}_{\alpha\beta}(p,m)=&-g_{\alpha\beta}+\frac{p_{\alpha}p_{\beta}}{m^2},
\end{align}
with four-momentum $p$ and mass $m$.

In the next subsection, we shall discuss the parameters used in our 
numerical analysis, which will be presented later in 
Section~\ref{Sec:Result}.

\subsection{The parameters} 

In order to numerically calculate the radiative decay widths of $X_2$, we should 
fix the values for the parameters used. The values for the meson masses are \cite{ParticleDataGroup:2022pth}
$$m_{D}=1867.25~ \text{MeV},~ m_{D^*}=2008.56 ~\text{MeV}, ~m_{X_2}=4014.3 ~ \text{MeV},$$
$$m_{J/\psi}=3096.90 ~\text{MeV},~ m_{\psi(2S)}=3686.10 ~\text{MeV},$$
where $m_D$ ($m_{D^*}$) is the average mass between the neutral and charged 
$D$ ($D^*$) mesons, and the $X_2$ mass is taken from Ref.~\cite{Belle:2021nuv}. The charm quark mass and the parameter related to the 
magnetic coupling are fixed by the partial electromagnetic widths for $D^{*0}\to 
\gamma D^0$ and $D^{*+}\to \gamma D^+$ \cite{Hu:2005gf}, 
\begin{align}
    \beta'^{-1}=379~ \text{MeV},~ m_c=1863~\text{MeV}.
    \label{Eq:parameter_magnitic}
\end{align}

The coupling constant for $X_2$ to the charged and neutral charmed mesons is 
extracted from the binding energy of $X_2$ \cite{Weinberg:1965zz,Baru:2003qq},
\begin{align}
    \chi^{0}_{\text{nr}}=\left\{\lambda^2\frac{16\pi}{\mu_{*0}}\sqrt{\frac{2E_B}{\mu_{*0}}}\left[1+{\cal O}(\sqrt{2\mu_{*0} E_B}r)\right]\right\}^{1/2},
\end{align}
where $E_B$ and $\mu_{*0}$ are the binding energy of $X_2$ relative to the $D^{*0}\bar D^{*0}$ threshold and the $D^{*0}\bar D^{*0}$ reduced mass, respectively. 
$r$ is identified with the range of forces, where $1/r\gg\sqrt{2 \mu_{*0} E_B}$ 
in the weak binding limit \cite{Baru:2021ldu}. 
We assume that the $X_2$ is a pure $D^*\bar D^*$ bound state, $\lambda^2=1$, and then we obtain the 
coupling constant $\chi_{\text{nr}}=1.3^{+1.0}_{-1.3} ~\text{GeV}^{-1/2}$, where the uncertainty is derived from the uncertainties of   $D^{*0}$ ($\bar D^{*0}$) and $X_2$ masses. 
Besides, $g$ and $g'$ denote the $J/\psi$ and $\psi(2S)$ couplings to the 
charmed mesons, respectively.

We evaluate the loop integrals in Fig. \ref{Fig:decay_X_2} by using the 
dimensional regularization method. In particular, we adopt the $\overline{\text{MS}}$ 
subtraction scheme.
As for the strength of the interaction corresponding to the counterterms, 
denoted by the $\lambda_r$ parameters,  following the idea in Ref.~\cite{Guo:2014taa}, we 
set to zero the contribution of the finite part of the counterterms in 
Eq.~\eqref{Eq:amp_conuterterm} and vary the energy scale in a large range, 1.5--7.0 GeV, in the UV divergent loop integrals. We define the ratios
\begin{align}
 r_{\chi}\equiv\left|\frac{\chi_{\text{nr}}}{\bar \chi_{\text{nr}}}\right|,\quad r'_{\chi}\equiv\left|\frac{\chi'_{\text{nr}}}{\bar \chi'_{\text{nr}}}\right|,\quad r_{g}\equiv\left|\frac{g_2}{g^0_2}\right|,\quad r'_{g}\equiv\left|\frac{g'_2}{g^0_2}\right|,
\label{Eq:ratio_coupling}
\end{align}
where $g_2$ ($g^{\prime}_2$) is the coupling constant between $J/\psi$ ($\psi(2S)$) and charmed mesons in Eq.~\eqref{Eq:Coupling_g2}. We take $\bar \chi_{\text{nr}}=1.3~\text{GeV}^{-1/2}$ and $g_2^0 = 2.0~\text{GeV}^{-3/2}$ with the latter
from the model-dependent estimates discussed in Refs.~\cite{Colangelo:2003sa,Guo:2010ak}.
The $X(3872)$ coupling to $D\bar D^*$ is denoted as $\chi'_\text{nr}$, and its benchmark value is set to $\bar \chi'_{\text{nr}}=0.97 ~\text{GeV}^{-1/2}$~\cite{Guo:2013zbw}.

In what follows, we present the numerical results for the $X_2$ radiative decays into $\gamma\psi$. In order to perform the numerical analysis, 
we have used the following Mathematica packages: FeynCalc~\cite{Shtabovenko:2020gxv}, 
FeynHelpers~\cite{Shtabovenko:2016whf}, and Package-X~\cite{Patel:2016fam}.

\section{Numerical results} 
\label{Sec:Result}

In Table~\ref{Tab:X_2_Decay}, we show our numerical results 
for the partial decays $X_2 \to \gamma \psi$ with $\psi=\psi(2S)$ 
and $J/\psi$ as given in Eq.~\eqref{Eq:X2decay}, along with 
the corresponding ratio $R_{X_2}$. These observables 
depend on the products among the quantities defined in Eq.~\eqref{Eq:ratio_coupling}. 
In order to fix those quantities, we have to make assumptions on the 
coupling constants $g_2$ and $g_2^{\prime}$ since their values are not 
well established in the literature. Besides, we have also 
to fix the contribution from the contact terms, that is, to fix the 
size of $\lambda_r$'s in Eq.~\eqref{Eq:amp_conuterterm}, which is unknown; 
however, as discussed Ref.~\cite{Guo:2014taa}, we may estimate their size by noticing that any change in 
$\mu$ must also change the counterterms accordingly so that the overall 
result does not depend on the renormalization scale $\mu$. Thus, we set $\lambda_r=0$ and
vary the scale $\mu$ within a large range from $1.5$ GeV up to $2 m_{X_2}$, 
with $m_{X_2}$ the $X_2$ mass, as mentioned above. In Table~\ref{Tab:X_2_Decay}, we can see 
such behavior in the partial decays under consideration by noticing 
their corresponding changes as $\mu$ varies. In other words, such variation 
in each partial decay width we are concerned with may be considered a measure 
of the size of $\lambda_r$'s.

\begin{table}[hbt!]
\caption{\label{Tab:X_2_Decay} Decay widths and their ratio $R$ for the process $X\rightarrow \gamma \psi$ with $X=X(3872), X_2$ and $\psi=J/\psi, \psi(2S)$. 
 $\Gamma'_{\psi}$ and $\Gamma_{\psi}$ denote the decay width for $X(3872)\to \psi\gamma$ and $X_2\to \psi\gamma$, respectively.
  The first row is the energy scale in the $\overline{\text{MS}}$ subtraction scheme,
$g_2$ ($g'_2$) is the coupling constant in Eq.~\eqref{Eq:Coupling_g2}, and  $r'_{\chi}$, $r_g$, and $r_g'$ are defined in Eq.~\eqref{Eq:ratio_coupling}.
In the last row, $N_{\text{min}}(X_2\to \gamma J/\psi)$ denotes the lower limit of the $X_2$ signal yield  in the $J/\psi \gamma\to \ell^+\ell^-\gamma$ ($\ell=e,\mu$) mode for the two-photon process at Belle. }
\renewcommand{\arraystretch}{1.2}
\begin{tabular*}{\columnwidth}{@{\extracolsep\fill}lcccc}
\hline\hline                                           
         $\mu$ (GeV) & 1.5 & 2.0   & 4.0 & 7.0 \\[3pt]        
\hline
     $\Gamma'_{J/\psi}$ (keV)  & 162 $(r'_{\chi}r_g)^2$  & 176 $(r'_{\chi}r_g)^2$   & 212 $(r'_{\chi}r_g)^2$   & 244 $(r'_{\chi}r_g)^2$ \\[3pt]
     $\Gamma'_{\psi(2S)}$ (keV)  & 17.5 $(r'_{\chi}r_g')^2$  & 18.4 $(r'_{\chi}r_g')^2$   & 20.8 $(r'_{\chi}r_g')^2$   & 22.7 $(r'_{\chi}r_g')^2$ \\[3pt]
     $\Gamma_{J/\psi}$ (keV)   & 139 $(r_{\chi}r_g)^2$   & 161 $(r_{\chi}r_g)^2$   & 224 $(r_{\chi}r_g)^2$   & 284 $(r_{\chi}r_g)^2$ \\[3pt]
     $\Gamma_{\psi(2S)}$ (keV)  & 25.0 $(r_{\chi}r_g')^2$  & 27.1 $(r_{\chi}r_g')^2$   & 32.7 $(r_{\chi}r_g')^2$   & 37.6 $(r_{\chi}r_g')^2$ \\[3pt]
     $R_{X(3872)}$  & 0.11 $(g_2'/g_2)^2$  & 0.10 $(g_2'/g_2)^2$   & 0.10 $(g_2'/g_2)^2$   & 0.09 $(g_2'/g_2)^2$ \\[3pt]
     $R_{X_2}$     & 0.18 $(g_2'/g_2)^2$ & 0.17 $(g_2'/g_2)^2$   & 0.15 $(g_2'/g_2)^2$   & 0.13 $(g_2'/g_2)^2$ \\[3pt]
     $R_{X_2}/R_{X(3872)}$ & 1.67 & 1.61  & 1.49    & 1.43 \\[3pt]
$N_{\text{min}}(X_2\to \gamma J/\psi)$  & 35  & 36  & 39   & 41 \\[3pt]
\hline\hline
\end{tabular*}
\end{table}

As for the ratio $R_{X_2}$, we set the finite part of the counterterm to 
zero and $R_{X_2}$ will depend only on the ratio 
$g_2^{\prime}/g_2$ at a given scale. In order to fix this latter ratio, we consider some 
model-dependent estimates. If  $g_2^{\prime} /g _2$ is assumed to equal to unity, 
for the $X(3872)$ case the $\gamma \psi(2S)$ channel is relatively suppressed~\cite{Guo:2014taa}. 
Similarly, in our case, by considering $g_2^{\prime} /g _2=1$, we 
find that the $\gamma \psi(2S)$ channel is also suppressed. In 
Ref.~\cite{Dong:2009uf}, the ratio was fixed to $g_2^{\prime} /g _2=1.67$ 
by modeling the couplings in a vector-meson dominance picture~\cite{Colangelo:2003sa}; in this case, the $\gamma \psi(2S)$ channel 
is still suppressed relatively to  the $\gamma J/\psi$ channel.
Alternatively, we can fix $g_2^{\prime} / g_2$ 
by using the upper limit for the ratio $R_{X(3872)}$ corresponding to the 
$X(3872) \to \gamma \psi(2S)$ and $\gamma J/\psi$, reported by BESIII 
in Ref.~\cite{BESIII:2020nbj}. The $X(3872)$ radiative decays are discussed 
in Ref.~\cite{Guo:2014taa} (see Appendix \ref{Sec:X_3872} which updates a couple of expressions in Ref.~\cite{Guo:2014taa}). The results for the $X(3872)$ state are 
also shown in Table~\ref{Tab:X_2_Decay}. Note that, in this case, 
the $R_{X(3872)}$ barely changes as we vary the scale $\mu$. Thus, we can 
choose one specific value for $\mu$, for instance, $\mu=1.5$ GeV, and 
then, by using the BESIII measurement for $R_{X(3872)}<0.59$, we obtain from 
Table~\ref{Tab:X_2_Decay}
\begin{equation}
    g_2^{\prime}/g_2 < 2.34. \label{eq:upper}
\end{equation}

\begin{figure}[tbp]
    \includegraphics[width=0.45\columnwidth]{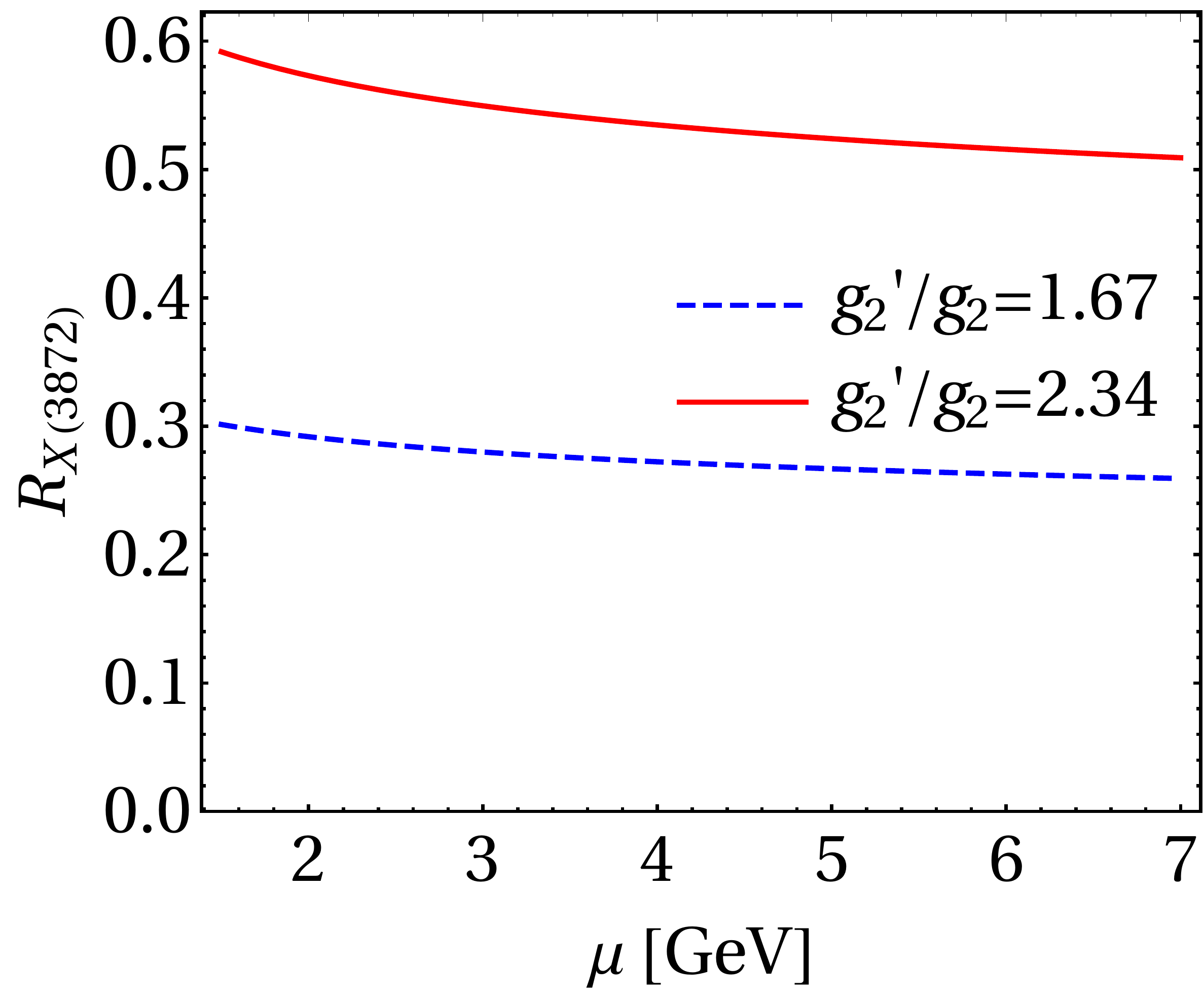} \hfill 
    \includegraphics[width=0.45\columnwidth]{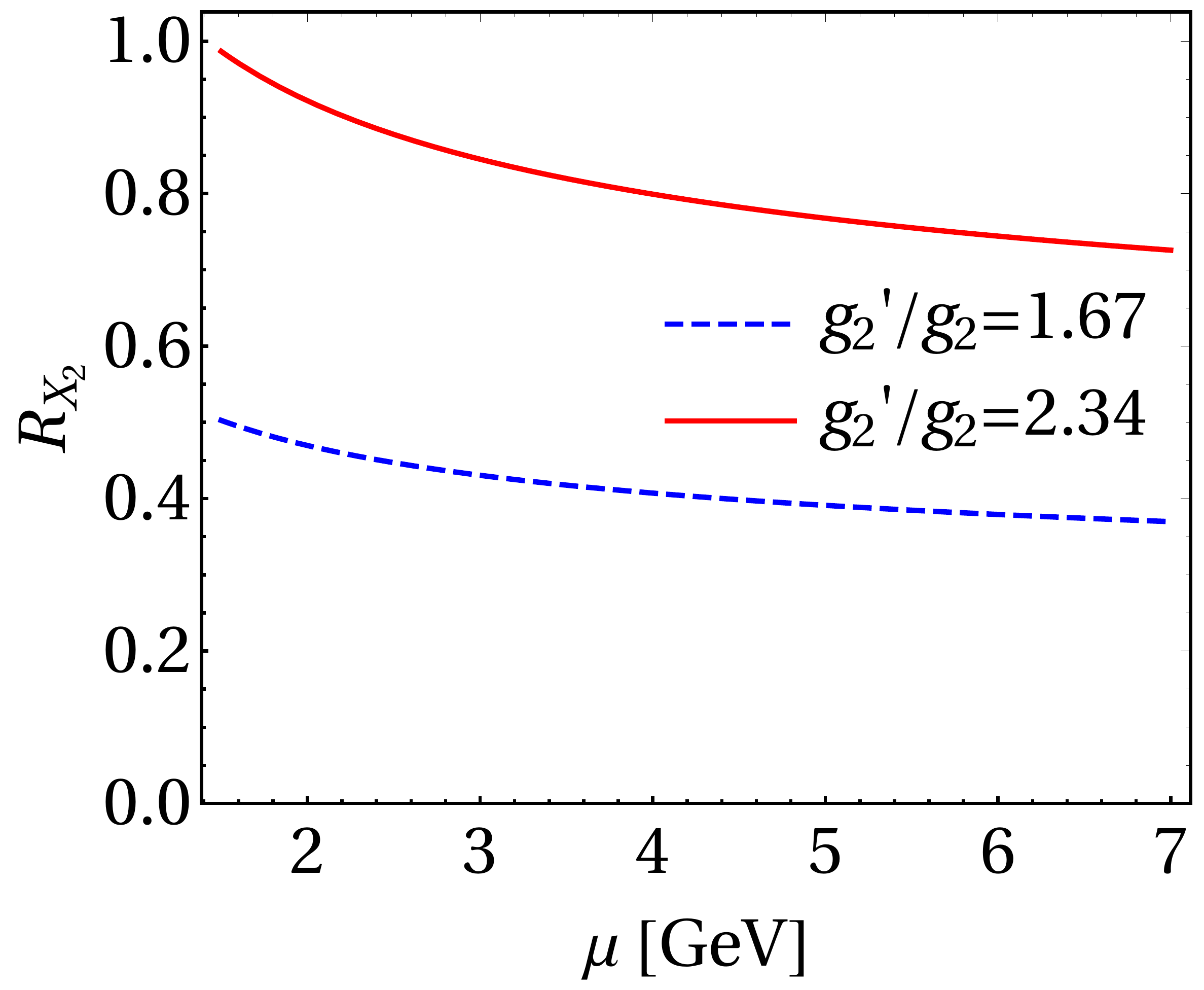}
    \caption{\label{Fig:ratio}~Scale dependence of the ratios $R_{X(3872)}$ and $R_{X_2}$, where $\mu$ is the energy scale in the $\overline{\text{MS}}$ 
    subtraction scheme of the dimensional regularization method used to regularize the 
    loops in Fig.~\ref{Fig:decay_X_2}. $g_2'/g_2=2.34$ is the upper bound given in Eq.~\eqref{eq:upper}, derived from the BESIII measurement $R_{X(3872)}<0.59$~\cite{BESIII:2020nbj}, and $g_2'/g_2=1.67$ is the model value used in Ref.~\cite{Dong:2009uf}.  }   
\end{figure}
As a matter of checking, in Fig.~\ref{Fig:ratio}, we show the plots for both 
ratios $R_{X(3872)}$ and $R_{X_2}$ as a function of the scale $\mu$. As we can 
see, within the range $1.5~\text{GeV} \leqslant \mu \leqslant 2 m_X$, those 
ratios are, to a good approximation, independent of $\mu$, as we expect. 

\begin{figure}[tb]
 \includegraphics[width=0.45\textwidth]{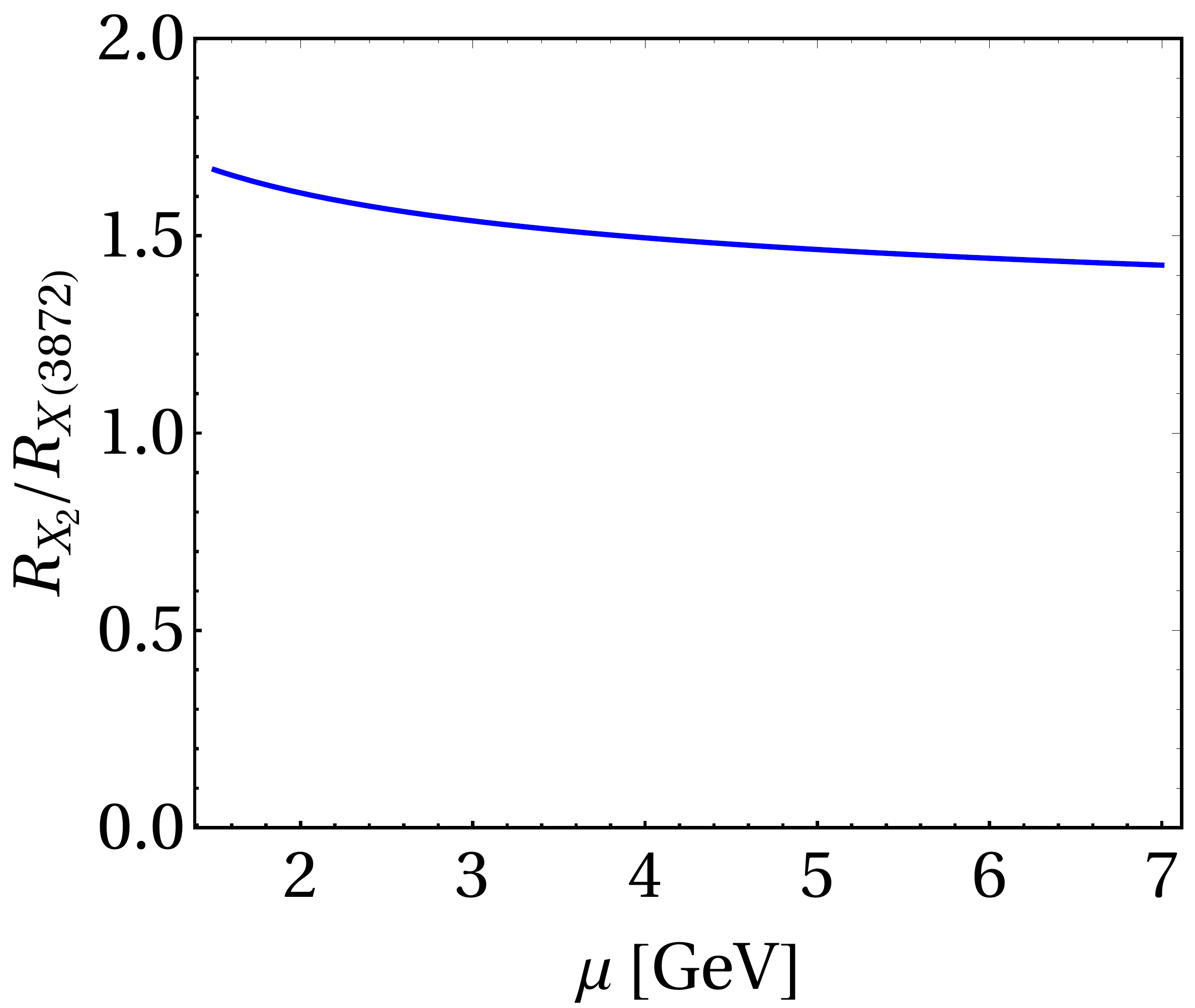}
 \caption{Scale dependence of the double ratio $R_{X_2}/R_{X(3872)}$.}\label{Fig:ratios_R}
\end{figure}

As discussed above, both ratios $R_{X(3872)}$ and $R_{X_2}$ depend on $g_2^{\prime}/g_2$ 
value, whose value is not fixed and model dependent. However, we can determine $R_{X_2}$ 
independently of the couplings $g_2^{\prime}$ and 
$g_2$, using the experimental information for $R_{X(3872)}<0.59$, measured by BESIII in 
Ref.~\cite{BESIII:2020nbj}, as an input. 
Specifically, in Fig.~\ref{Fig:ratios_R} we show the double ratio $R_{X_2}/R_{X(3872)}$ as a 
function of $\mu$. Note that this observable only slightly decreases from $1.67$ to 
$1.43$ as the renormalization scale $\mu$ varies within the large range $[1.5~\text{GeV}, 2\,m_{X_2}]$. This flat variation indicates that we can set any value within that range 
for $R_{X_2}/R_{X(3872)}$, and then use the upper boundary $R_{X(3872)} < 0.59$ provided by 
BESIII such that we obtain an upper limit for $R_{X_2}$: 
\begin{equation}
    R_{X_2}\lesssim 1.0.
\end{equation}

Next, we shall estimate the number of events of $X_2$ that can be collected in the $\gamma J/\psi$ final state of the two-photon process at the Belle 
experiment. The 
signal yield of $X_2$ in such a process is given by 
\begin{equation}
N(X_2 \rightarrow \gamma \psi)=\Gamma_{\gamma \gamma} 
\operatorname{Br}\left[X_2\right] \operatorname{Br}[\psi] \epsilon F(\sqrt{s}, J) L_{\text {tot }},
\end{equation}
where $\Gamma_{\gamma \gamma}$ is the two-photon decay width of $X_2$, 
$\textrm{Br}[X_2]$ is the branching fraction for $X_2 \to \gamma 
\psi(\psi=J / \psi, \psi(2 S))$, $\epsilon$ is the efficiency, and 
$L_{\text {tot }}$ is the total integrated luminosity of the Belle data 
sample. In addition, 
for the $\psi(2S)$ reconstructed from $J/\psi\pi^+\pi^-$ as done in Ref.~\cite{Belle:2021nuv} and $J/\psi$ from the $\ell^+\ell^-$ lepton pairs ($\ell=e, \mu$), one has
$\textrm{Br}[\psi(2 S)]=\textrm{Br}\left[\psi(2 S) \to \pi^{+} \pi^{-} 
J / \psi\right] \textrm{Br}\left[J / \psi \rightarrow \ell^{+} \ell^{-}\right]$ 
and $\textrm{Br}[J / \psi]=$ $\textrm{Br}\left[J / \psi \to \ell^{+} \ell^{-}\right]$. 
The factor $F(\sqrt{s}, J)$ 
is related to the two-photon luminosity function $L_{\gamma \gamma}$~\cite{Uehara:1996bgt,Belle:2021nuv}
\begin{equation}
    F(\sqrt{s}, J)=4 \pi^2(2 J+1) L_{\gamma \gamma}(\sqrt{s}) / s,
\end{equation}
with the effective energy of the two-photon collision $\sqrt{s}$ 
and the spin $J$ for $X_2$. Since the two-photon decay width $\Gamma_{\gamma \gamma}$, the efficiency $\epsilon$, the total 
integrated luminosity $L_{\text {tot }}$ and the factor $F(\sqrt{s}, J)$ 
are the same for $N\left(X_2 \to \gamma 
J / \psi\right)$ and $N\left(X_2 \to \gamma \psi(2 S)\right)$, the signal 
yield of $X_2$ in the $\gamma J / \psi$ with $J/\psi\to \ell^+\ell^-$ is 
given by 
\begin{equation}
N(X_2 \to \gamma J/\psi)=\frac{N(X_2 \to \gamma \psi(2 S))}{R_{X_2} 
\textrm{Br}[\psi(2 S) \to \pi^{+} \pi^{-} J / \psi]}\, .
\end{equation}
The signal yield of $X_2$ is $19 \pm 7$ in Ref.~\cite{Belle:2021nuv}, and the branching fraction for $\psi(2 S) 
\to \pi^{+} \pi^{-} J / \psi$ is $(34.68 \pm 0.30) \%$~\cite{ParticleDataGroup:2022pth}. As discussed above, for $\mu=1.5$~GeV 
we have $R_{X_2}\lesssim1.0$, the yield of 
$X_2$ is 
\begin{equation}
    N\left(X_2 \rightarrow \gamma J / \psi\right)\gtrsim 35.
\end{equation}
Therefore, we estimate that the signal yield of $X_2$ observed in the 
$\gamma J / \psi$ invariant mass distribution is at least about 35 for the 
two-photon collision at Belle. Besides, the minimal yields of $X_2$ 
estimated with other energy scales are listed in Table~\ref{Tab:X_2_Decay}.
The prediction can be checked with the Belle data.

\section{Summary}
\label{Sec:Summary}

By assuming the existence of a tensor state $X_2$, 
that, according to HQSS, is a spin partner 
of the $X(3872)$ state in the hadronic molecular model, as predicted in Refs.~\cite{Nieves:2012tt,Tornqvist:1993ng,Guo:2013sya}, 
we have evaluated its radiative decays into $\gamma\, \psi(2S)$ 
and $\gamma\,J/\psi$ channels. 
Although with low statistics, a candidate of such a state 
was recently observed by the Belle Collaboration in Ref.~\cite{Belle:2021nuv}, 
with mass and width in accordance with the corresponding values 
predicted in Refs.~\cite{Nieves:2012tt,Tornqvist:1993ng,Guo:2013sya}.

In particular, in our case, the decays we are concerned with 
proceed through hadronic loops with charmed mesons as intermediate 
particles. These loops are UV divergent, requesting an introduction of 
additional counterterm amplitude, in which the strength of that contact 
interaction absorbs the infinities after renormalization. However, with 
the available theoretical information, it is impossible to determine 
the contribution from the counterterms precisely. Notwithstanding, we 
estimated their contributions by varying the renormalization scale, as done in 
Ref.~\cite{Guo:2014taa} for the $X(3872)$ case, such that the changes 
in the partial decay widths encode the size of the counterterm.

Moreover, we have used effective Lagrangians to describe the 
$X_2$ couplings to the charmed $D^*$ and $\bar{D}^*$ mesons, 
and the couplings of those mesons with the charmonia $\psi(2S)$ 
and $J/\psi$. Since these latter set of couplings are not well-determined 
in literature, we have written the partial decays $X_2\to 
\gamma\,\psi(2S)$ and $X_2\to \gamma\,J/\psi$ in terms of ratios 
involving those couplings, that allows us to draw conclusions 
based on the relations among them. 

According to our findings, for values of $g_2^{\prime}/g_2$ 
close to one, we always find suppression of the $\gamma\,\psi(2S)$ 
channel against the $\gamma\,J/\psi$ one. On the other hand, 
as $g_2^{\prime}/g_2$ increases the ratios $R_{X(3872)}$ and $R_{X_2}$ increase accordingly. 
We have fixed the range of $g_2^{\prime}/g_2$ to be $<2.34$ using input from the BESIII measurement of $R_{X(3872)}<0.59$.
Consequently, we found $R_{X_2}\lesssim1.0$.
As a matter 
of comparison, $R_{X_2}$ estimated using the quark model assuming the $X_2$ to be the $\chi_{c2}(2P)$ meson leads to a value about 4, as displayed in Table~\ref{Tab:charmonium}, significantly larger than the one we obtained in the hadronic molecular picture. 
In principle, future experimental measurements of the observable 
$R_{X_2}$ may shed light on the internal structure of $X_2$.

Finally, we predict the signal yield of $X_2$ in the $\gamma\,J/\psi$ spectrum 
from the two-photon process based on our findings for $R_{X_2}$, and also the 
yields of $X_2$ in the $\gamma\,\gamma\to \gamma \, \psi(2S)$ reaction reported 
by the Belle Collaboration~\cite{Belle:2021nuv}. As a result, we expect that the yield of $X_2$ 
in the $\gamma\,J/\psi\to \gamma \ell^+\ell^-$ final states $(\ell = e,\mu)$  should be at least about $35$, that is larger than the number of events 
observed by the Belle collaboration in Ref.~\cite{Belle:2021nuv}.

\begin{acknowledgments}
    We would like to thank Alexey Nefediev for valuable discussions. This work is partly supported by the Chinese Academy of Sciences under Grant 
    No.~XDB34030000; by the National Natural Science Foundation of China (NSFC) under 
    Grants No.~12125507,  No.~11835015, and No.~12047503; and by the NSFC and the Deutsche Forschungsgemeinschaft 
    (DFG) through the funds provided to the Sino-German
Collaborative  Research Center ``Symmetries and the Emergence of Structure in QCD'' (NSFC Grant No. 12070131001, DFG Project-ID 196253076 - TRR110).
\end{acknowledgments}

\appendix

\section{$X(3872)$ radiative decays into $\gamma\,\psi(2S)$ and $\gamma\,J/\psi$ channels}
\label{Sec:X_3872}

The $X(3872)$ radiative decays into the $\gamma\,\psi(2S)$ and $\gamma\,J/\psi$ channels 
have been evaluated in Ref.~\cite{Guo:2014taa}, adopting a molecular picture as the 
quark configuration for the $X(3872)$ state. The findings reported in Ref.~\cite{Guo:2014taa} 
are in line with the experimental ratio $R_{X(3872)}$ measured by the BESIII Collaboration~\cite{BESIII:2020nbj}. In particular, it is stressed in Ref.~\cite{Guo:2014taa} that 
those specific decays take place through hadron loops involving charmed mesons plus short-distance contributions in the form of a counterterm.
For pure hadronic molecular states the hadron loops are the leading order contribution 
and play an important role in such processes. 
However, the charged conjugated diagrams in the loops are not considered in Ref.~\cite{Guo:2014taa}; these contributions will lead to a factor of 2 for all the loop contributions but does not affect the ratio $R_{X(3872)}$ which was the main concern in Ref.~\cite{Guo:2014taa}.
In addition, $p^\alpha$ in Eq.~(24) and $p^\gamma$ in Eq.~(28) in Ref.~\cite{Guo:2014taa} should be changed to $k^\alpha$ and $(k-p)^\gamma$, respectively, which has also been noticed in Ref.~\cite{Baru:2017fgv}.
That is because the four-velocity of heavy mesons with respect to the magnetic 
vertices in Eqs.~\eqref{Eq:vertex_D1_D_gamma_M} and 
\eqref{Eq:vertex_D1B_DB_gamma_M} are related to the charmed-meson momentum inside the loop, instead of the $X(3872)$ four-velocity. 
The updated expressions of Eqs.(24)-(29) in Ref.~\cite{Guo:2014taa} for the individual contributions to the process $X_{\sigma}(p)\to \gamma_{\lambda}(q)\psi_{\mu}(p-q)$ are (we use the same notation as in Ref.~\cite{Guo:2014taa})
\begin{align}
    J_{\mu\nu\lambda}^{(a)m}(k)=&\,\frac{2}{3}m \left(\beta+\frac{4}{m_c}\right)\epsilon_{\nu\lambda\alpha\beta}k^{\alpha}q^{\beta}\frac{(2k-p-q)_{\mu}}{(k-q)^2-m^2},\nonumber\\[3pt]
     J_{\mu\nu\lambda}^{(b)e}(k)=&\,4\epsilon_{\mu\rho\alpha\beta}\frac{(k-p)^{\alpha}(k-q)^{\beta}}{(k-q)^2-m_{*}^2} \left[(2k-q)_{\lambda} g^{\rho}_{\nu}-(k-q)_{\nu} g^{\rho}_{\lambda}-k^{\rho}g_{\nu\lambda}\right],\nonumber\\[3pt]
    J_{\mu\nu\lambda}^{(b)m}(k)=&\,\frac{4}{3}m_{*} \left(\beta-\frac{4}{m_c}\right)\epsilon_{\mu\rho\alpha\beta}\frac{(k-p)^{\alpha}(k-q)^{\beta}}{(k-q)^2-m_{*}^2}\left[q_{\nu} g^{\rho}_{\lambda}-q^{\rho} g_{\nu\lambda}\right],\nonumber\\[3pt]
    J_{\mu\nu\lambda}^{(c)e}(k)=&\,4\epsilon_{\mu\nu\alpha\beta}(k-p+q)^{\alpha}k^{\beta}\frac{(2k-2p+q)_{\lambda}}{(k-p+q)^2-m^2}, \nonumber\\[3pt]
    J_{\mu\nu\lambda}^{(d)m}(k)=&\,\frac{2}{3}m_{*} \left(\beta+\frac{4}{m_c}\right)
    \left[(2k-p+q)_{\mu}g_{\beta\nu}-(2k-p+q)_{\beta}g_{\mu\nu}-(2k-p+q)_{\nu}g_{\beta\mu}\right]\frac{\epsilon_{\alpha\lambda\gamma\delta}(k-p)^{\gamma}q^{\delta}}{(k-p+q)^2-m_{*}^2}\nonumber\\[3pt]
    &\left(-g^{\alpha\beta}+\frac{(k-p+q)^{\alpha}(k-p+q)^{\beta}}{m_{*}^2}\right),\nonumber\\[3pt]
    J_{\mu\nu\lambda}^{(e)e}(k)=&-4\epsilon_{\mu\nu\lambda\alpha}p^{\alpha},
    \label{Eq:amplitude_X_3872}
\end{align}
where $m$ and $m_{*}$ are the masses of $D$ and 
$D^*$ mesons, respectively, and the magnetic coupling parameter $\beta$ is the 
$\beta'$ in Eq.~\eqref{Eq:parameter_magnitic}.

\begin{table}[bt]
\caption{\label{Tab:X_3872} Decay widths $X(3872)\to \gamma\psi$ 
with $\psi=J/\psi, \psi(2S)$ and their ratio $R_{X(3872)}$. The 
second row displays the results for $X(3872)\to \gamma\psi$ with 
$\psi=J/\psi,~\psi(2S)$ calculated in Ref. \cite{Guo:2014taa}, while 
the updated results are shown in the third row.
}
\renewcommand{\arraystretch}{1.2}
\begin{tabular*}{\columnwidth}{@{\extracolsep\fill}lcccc}
\hline\hline                                           
        & $\mu$    & $\Gamma'_{J/\psi}$ [keV]   & $\Gamma'_{\psi(2S)}$ [keV] & $R_{X(3872)}$  \\[5pt]        
\hline
  Ref.~\cite{Guo:2014taa} & $m_{X(3872)}$ & 23.5 $(r'_{\chi}r_g)^2$  
   & 4.9 $(r'_{\chi}r'_g)^2$   & 0.21 $(g_2'/g_2)^2$  \\[5pt]
  Updated & $m_{X(3872)}$  & 211 $(r'_{\chi}r_g)^2$  
  & 20.6 $(r'_{\chi}r'_g)^2$   & 0.10 $(g_2'/g_2)^2$ \\[5pt]
\hline\hline
\end{tabular*}
\end{table}

The updated  numerical results for the partial decay widths of 
$X(3872)\to \gamma\,\psi(2S)$ and $\gamma\,J/\psi$ are shown in 
Table~\ref{Tab:X_3872}, where we have kept the results from Ref.~\cite{Guo:2014taa} for a comparison. As can be seen, they are much larger than 
the ones in Ref.~\cite{Guo:2014taa}. As for the ratio $R_{X(3872)}$, the new result is about half of the previous one.
Nevertheless, the conclusion in Ref.~\cite{Guo:2014taa}, that is the hadronic molecular picture of the $X(3872)$ is compatible with the measured  ratio $R_{X(3872)}$, is not altered.

\bibliography{radiative_decay_X2_ref.bib}

\end{document}